\documentclass{aa}
\usepackage{graphicx}
\usepackage{txfonts}
\usepackage{natbib}
\bibpunct{(}{)}{;}{a}{}{,} % to follow the A&A style

\begin{document}

\title{An XMM-Newton view of the X-ray flat radio-quiet quasar PG\,1416$-$129}
\subtitle{}
\author{D.\ Porquet\inst{1}
\and J.~N.\ Reeves\inst{2,3}
\and A.\ Markowitz\inst{2,4}
\and T.~J.\ Turner\inst{2,5}
\and L.\ Miller\inst{6}
\and K.\ Nandra\inst{7}}

\offprints{D. Porquet}

\institute{
Max-Plank-Institut f\"{u}r extraterrestrische Physik, Postfach 1312,
D-85741, Garching, Germany\\
\email{dporquet@mpe.mpg.de}
\and 
Laboratory for High Energy Astrophysics, Code 662, 
NASA Goddard Space Flight Center, Greenbelt, MD 20771, USA
\and Dept. of Physics and Astronomy, Johns Hopkins University, 3400 N.\ 
Charles Street, Baltimore, MD 21218, USA
\and NASA Postdoc Program Associate
\and Dept. of Physics, University of Maryland Baltimore County, 
1000 Hilltop Circle, Baltimore, MD 21250, USA
\and Dept. of Physics, University of Oxford, Denys Wilkinson Building, Keble Road, Oxford OX1 3RH, UK
\and Astrophysics Group, Imperial College London, Blackett Laboratory,
Prince Consort Road, London SW7 2AW, UK}

\date{Received  / Accepted }

\abstract
%context heading (optional), leave it empty if necessary 
{}
{The radio-quiet quasar PG\,1416$-$129 ($z=0.129$) exhibits 
atypical optical and
  X-ray properties. Between 1990 and 2000, in response to 
its optical continuum decrease, the ``classical'' broad component of 
H$\beta$ almost completely disappeared, with a factor of 10 
decrease in the line flux. In addition, the width of the broad component of
the H$\beta$ line decreased significantly from 4000 km s$^{-1}$ 
to 1450 km s$^{-1}$.  
In the X-ray band, this object was observed by Ginga 
in 1988 to have the hardest quasar 
photon index, with $\Gamma$=1.1$\pm$0.1. 
We present an {\sl XMM-Newton}/EPIC observation of PG\,1416$-$129
 performed in July 2004. 
}
{We analyze the time-averaged pn spectrum of this quasar, as well
  as perform time-resolved spectroscopy.}
{We find that during the present XMM-Newton observation, PG\,1416$-$129 
still has a rather hard photon index,  both in the soft (0.2--2\,keV)
and hard (2--12 keV) energy ranges, compared to radio-quiet quasars
(BLS1 and NLS1) but compatible with the photon index value found for
radio-loud quasars. This object also shows long-term luminosity 
 variability over 16 years by a factor of three with a variation of
 photon index from $\sim$1.2 to $\sim$ 1.8. 
In the soft energy band (0.2--2\,keV), we found a very weak soft X-ray
 excess compared to other RQ quasars. 
 The whole time averaged spectrum is fit very well either by 
X-ray ionized reflection from the accretion disk surface, by a warm
absorber-emitter plus power-law, or by a smeared
absorption/emission from a relativistic outflow. 
While no constant narrow Fe\,K line at 6.4\,keV is observed, we find
the possible presence of two non-simultaneous transient iron lines:  
a redshifted narrow iron line at about 5.5\,keV (96.4$\%$
confidence level according to multi-trial Monte-Carlo simulations) at the beginning 
of this observation and the appearance of a line at 6.3--6.4\,keV
(99.1$\%$ c.l.) 
at the end of the observation. These variable lines could be generated 
by discrete hot-spots on the accretion disk surface. }
{}

\titlerunning{XMM-Newton view of PG\,1416$-$129}
\authorrunning{D.\ Porquet et al.}

\maketitle

\section{Introduction}

\object{PG\,1416$-$129} ({\it z} $= 0.129$) was one of the three brightest low-redshift 
(z$<$0.2) radio-quiet quasars observed during the ASCA mission \citep{RT00}. 
It has a high bolometric luminosity, 10$^{46}$\,erg\,s$^{-1}$ 
\citep{Vest02}, and a black hole mass of about 
3--6 $\times$ 10$^{8}$\,$M_{\sun}$ \citep{Vest02, Hao05}.
This object has an accretion rate with respect to Eddington of about 
0.12 \citep{Hao05}.
It has been associated with the 2MASS object at  
$\alpha_{J2000}$=14$^{\rm h}$19$^{\rm m}$03\fs81, 
 $\delta_{J2000}$=$-$13$^{\rm \circ}$10\arcmin44\farcs7 \citep{BH01}. 
It has strong, broad permitted optical emission lines:
 $FWHM$(H${\beta}$)= 6110\,km\,s$^{-1}$ \citep{BG92}, but 
 the measurement of this width was contaminated by the presence of
  a very broad component, as shown by \citet{Sulentic00}, 
who measured FWHM(H${\beta}$)= 4000\,km\,s$^{-1}$.  
 This value of the line width led to it being described as
 a Broad-Line (BL) AGN (FWHM(H${\beta})>$2000\,\,km\,s$^{-1}$). 
However, observations of this broad-line quasar have 
revealed a significant drop-off in its continuum luminosity by 
about a factor of four at 4500\,\AA~  
 between 1990 and 2000 \citep{Sulentic00}. In response to 
this continuum decrease, the ``classical'' broad component of 
H${\beta}$ has almost completely disappeared, with a factor of 10 
decrease in the line flux \citep{Sulentic00}. 
Additionally, the width of the broad component of the H${\beta}$ line 
decreased significantly over those 10 years, from 4000\,km\,s$^{-1}$ 
to 1450\,km\,s$^{-1}$.  \\
In the X-ray band, PG\,1416$-$129 shows peculiar behavior
compared to other low-redshift quasars. A spectrum obtained  
by {\sl Ginga} in 1988 showed the hardest spectral index ($\Gamma$=1.1$\pm$0.1) 
of all known quasars in the 2--20\,keV energy range \citep{Williams92}.
The {\sl Compton Gamma-ray Observatory}/OSSE observations in 1994 
showed that PG\,1416$-$129 has a very steep
spectrum ($\Gamma$=3.2$\pm$0.5) above 50\,keV,
 and consequently the most dramatic cut-off between 50\,keV 
and 100\,keV \citep{Staubert96}. However, from an {\sl ASCA} observation in 1994, 
 \citet{RT00} found a photon index for this object of $\Gamma$=1.78$\pm$0.02, 
which is closer to the values found in most radio-quiet low-redshift quasars
 (e.g., \citealt{P04a}) than the very hard photon index
 measured previously with {\sl Ginga}. 
 In addition, during this {\sl ASCA} observation, a possible
  Fe\,K$\alpha$ line was detected at $E$=6.54$^{+0.16}_{-0.18}$\,keV
  ($EW$=140$\pm$75\,eV) with an  F-test probability of 98.3$\%$.  
 This object also shows long-term flux variability as reported in
Table~\ref{tab:flux}. In this paper we report the analysis of the {\it XMM-Newton} 
observation of PG\,1416$-$129 performed in July 2004. In addition to the previous X-ray 
observations already published and the present {\sl XMM-Newton}
observation, we also show the currently unpublished 1998 {\sl RXTE} data.
Details of the {\sl RXTE}  data analysis are reported in Appendix 
\ref{sec:rxte}.  \\
%--------------------- TABLE ---------------------
\begin{table}[!Ht]
\caption{Summary of the power-law index and luminosities
 ($\times$ 10$^{44}$\,erg\,s$^{-1}$) observed by
  different X-ray satellites in the 2--10\,keV energy range over 16
  years (5 observations). 
For a direct comparison, the luminosities have been re-calculated when needed 
assuming H$_{\rm 0}$=75\,km\,s$^{-1}$\,Mpc$^{-1}$ and q$_{0}$=0.5. 
The references (column 5) correspond to: RT00 \citep{RT00}, LT97 \citep{LT97}.} 
\begin{tabular}{llllllllll}
\hline
\hline
\noalign {\smallskip}                       
Satellites &  \multicolumn{1}{l}{Obs date} & \multicolumn{1}{c}{$\Gamma$} &  Lum   &   Ref \\
           &  {\tiny (dd/mm/yyyy)}&  \\
\noalign {\smallskip}                       
\hline
\noalign {\smallskip}                       
Ginga &  02/02/1988  & 1.18$\pm$0.06 & 1.4 & LT97\\
\noalign {\smallskip}                       
Ginga &  21/09/1991  & 1.48$\pm$0.13  & 1.2 & LT97\\
\noalign {\smallskip}                       
ASCA  &  29/07/1994 & 1.78$\pm$0.02    &3.0  & RT00 \\
\noalign {\smallskip}                       
RXTE  &   21/08/1998 &    1.17$\pm$0.15 &1.0 & This work \\
\noalign {\smallskip}                       
XMM-Newton & 14/07/2004 & 1.56$\pm$0.04   & 0.98 & This work \\ 
\hline
\hline
\end{tabular}
\label{tab:flux}
\end{table}

\section[]{XMM-Newton observation and data reduction}

{\sl XMM-Newton} observed PG 1416$-$129 
 on July 14, 2004 (ID\,0203770201; orbit 842; exposure time $\sim$\,50\,ks). 
The EPIC$-$MOS cameras \citep{Turner01}  
 operated in the Small Window mode, 
while the EPIC$-$pn camera \citep{Str01} 
was operated in the standard Full Frame mode. 
The data were re-processed and cleaned (net pn time exposure
$\sim$45ks) using the {\it XMM-Newton} {\sc SAS version 6.5.0} ~(Science
Analysis Software) package. Since the effect of pile-up was negligible,
 X-ray events corresponding to patterns
 0--12 and 0--4 events (single and double pixels) 
were selected for the MOS and pn, respectively. Only good X-ray events
(with FLAG=0) were included. 
The low-energy cutoff was set to 300 and 200\,eV for MOS and pn, respectively. 
In this article we present the pn results only, 
since the pn-CCD have a better sensitivity over a broader energy range
 (0.2--12\,keV) compared to the MOS CCDs. 
However, we verified that similar results were obtained using the MOS data. 
The pn source spectra were extracted 
using a circular region of diameter 30$^{\prime\prime}$ centered on
 the source position. 
PG\,1416$-$129 is by far the brightest X-ray source in this
30$^{\prime}$ EPIC field-of-view. 
Background spectra were taken from a box region
 (7.3$^{\prime}$ $\times$ 4$^{\prime}$) 
on the same CCD as the source (excluding X-ray point sources). 
The {\sc xspec v11.3} software package \citep{Arnaud96} was used for spectral 
analysis of the background-subtracted spectrum, 
using the response matrices and ancillary files derived from the SAS tasks 
{\sc rmfgen} and {\sc arfgen}. 
The pn spectrum was binned to give a minimum of 50 counts per bin.\\
\indent  The signal to noise ratio was not sufficient 
for reliable RGS data analysis, and the OM was blocked during this
observation.

\section[]{{\sl XMM-Newton} spectral analysis}\label{sec:spectra}

In all fits, we included both the Galactic column density 
($N^{\rm Gal}_{\rm H}$=6.80 $\times 10^{20}$\,cm$^{-2}$, 
 obtained from the {\sc coldens} program using the compilations of
 \citealt{DL90}), and a possible  
additional absorption component located at the quasar redshift
 ($N^{\rm int}_{\rm H}$). 
Note that all fit parameters are given in the quasar's rest frame, 
with values of H$_{\rm 0}$=75\,km\,s$^{-1}$\,Mpc$^{-1}$ 
and q$_{\rm 0}$=0.5 assumed throughout. 
The errors quoted correspond to 90$\%$ confidence ranges for one
 interesting parameter ($\Delta \chi^{2}$=2.71).
 Abundances are those of \cite*{An89}. 
In the following, we use the updated cross-sections for X-ray absorption by 
the interstellar medium ({\sc tbabs} in {\sc XSPEC}) from \cite*{Wilms00}.  

\begin{figure}[!Ht]
\resizebox{\hsize}{!}{\rotatebox{-90}{\includegraphics{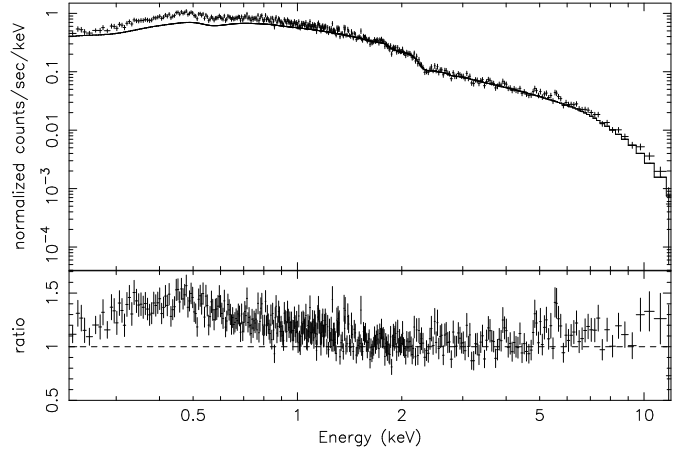}}}
\caption{The pn spectrum of PG\,1416$-$129 (in the observed frame).
A power-law has been fit to the 2--5~keV data
and extrapolated to lower and higher energies.
A weak soft X-ray positive residual is seen between 0.3--1\,keV, as well as
 the presence of a Fe\,K$\alpha$ complex between about 4.8--5.7\,keV 
($\sim$5.4--6.4\,keV in the quasar frame).
For presentation only, the data have been re-binned into groups of 10 bins,
after a grouping of a minimum of 20 counts per bin is used for the fit.
}
\label{fig:spectrum}
\end{figure}

\subsection{The time averaged spectrum}\label{sec:look}

We found an average pn count rate in the 0.2--12\,keV energy range 
 of 1.36$\pm$0.01\,cts\,s$^{-1}$.  
First, fitting the overall 0.2--12\,keV pn spectrum 
with a single absorbed power-law model, a moderately acceptable fit 
was found ($\chi^{2}_{\rm red}$= 1.18 for 979 d.o.f.), 
with $\Gamma$=1.75$\pm$0.01. 
To characterize the hard X-ray continuum, we fit an absorbed 
power-law model over the 2--5\,keV energy range, where the spectrum should
be relatively unaffected by the presence 
of a broad soft excess, a Warm Absorber-Emitter 
medium, an Fe\,K${\alpha}$ emission line, 
and a contribution above 8\,keV from 
a high energy Compton reflection hump.
In this energy range, the data are well fit by a single power-law model
 with  $\Gamma$= 1.61$\pm$0.06 ($\chi^{2}$/d.o.f.=346.2/377).  
This power-law index is similar to values found over the same energy range 
for radio-loud quasars 
 ($<$$\Gamma$=1.74$>$ with a standard deviation of 0.03; \citealt{P04a}),
but harder than the values commonly seen in radio-quiet quasars:
 $<$$\Gamma$=1.90$>$ (with a standard deviation of 0.27) and 
 $<$$\Gamma$=2.37$>$ (with a standard deviation of 0.11) for 
Broad Line quasars and Narrow Line quasars, respectively
(\citealt{P04a}). 
 A similar conclusion is found when comparing with the average 
  power law index 
measured in radio-quiet quasars above 2\,keV 
($<$$\Gamma_{2-10 \rm keV}$$>$= 2 with a dispersion of 0.25, 
\citealt{George00}; $<$$\Gamma_{2-12 \rm keV}$$>$=1.89$\pm$0.11,
\citealt{Pi05}).

Figure~\ref{fig:spectrum} displays the spectrum when the power-law model is
extrapolated over the 0.2--12\,keV broad band energy range. 
A weak positive residual is seen below 1\,keV 
due to the presence of a soft X-ray excess. 
 In addition, a Fe\,K$\alpha$ complex between about 4.8--5.7\,keV 
($\sim$5.4--6.4\,keV in the quasar frame) seems to be present.

%--------------------- TABLE ---------------------
\begin{table*}[!Ht]
\caption{Best-fit spectral parameters of the time-averaged pn spectrum 
  in the 2--12\,keV energy range
 with an absorbed (Galactic, $N_{\rm H}$=6.8$\times$10$^{20}$\,cm$^{-2}$) 
power-law (PL) component plus a line profile: 
 zgauss: Gaussian profile; and {\sc diskline} and {\sc laor}: line profile
emitted by a relativistic accretion disk for a non-rotating black hole
\citep{Fabian89} and  a maximally rotating black hole \citep{Laor91}, respectively. 
The line fluxes are expressed in 10$^{-6}$\,photons\,cm$^{-2}$\,s$^{-1}$. 
 We assume an emissivity index $q$ equal to $-$2.
 (a): $R_{\rm in}$=6\,$R_{\rm g}$ and $R_{\rm out}$=1000\,$R_{\rm g}$.
 (b): $R_{\rm in}$=1.235\,$R_{\rm g}$ and $R_{\rm out}$=400\,$R_{\rm g}$.}
\begin{center}
\begin{tabular}{llllllllll}
\hline
\hline
\noalign {\smallskip}                       
{\small Model}      &  \multicolumn{1}{c}{\small $\Gamma$} &\multicolumn{5}{c}{\small Line parameters}&{\small $\chi^{2}$/d.o.f.} &   F-test   \\
\noalign {\smallskip}                       
                   &                    &  \multicolumn{1}{c}{$E$ (keV)}    &  \multicolumn{1}{c}{$\sigma$ (keV)} &  \multicolumn{1}{c}{$\theta$ (deg)} & \multicolumn{1}{c}{Flux}  & \multicolumn{1}{c}{$EW$ (eV)} \\
\noalign {\smallskip}                       
\hline
\noalign {\smallskip}                       
{\small PL }       & {\small 1.56$\pm$0.04}&   \multicolumn{1}{c}{ -- } &   \multicolumn{1}{c}{ -- } &   \multicolumn{1}{c}{ -- } &   \multicolumn{1}{c}{ -- }   &    \multicolumn{1}{c}{ -- }       &     {\small 634.0/621} & \multicolumn{1}{c}{ -- }\\ 
\noalign {\smallskip}                       
\hline
\noalign {\smallskip}                       
{\small PL + zgauss}       & {\small 1.57$\pm$0.04} & {\small 6.4 (f)} &  0.01 (f)  &    \multicolumn{1}{c}{ -- }  &  {\small 2.2$\pm$1.2}   &  {\small 54$\pm$29}     &         {\small 624.5/620}&   99.78$\%$ \\ 
\noalign {\smallskip}                       
\hline
\noalign {\smallskip}                       
{\small PL + zgauss}       & {\small 1.57$\pm$0.04} & {\small 6.33$\pm$0.05 } & {\small $<$0.10} &    \multicolumn{1}{c}{ -- }  &  {\small 2.9$\pm$1.4} &  {\small 69$^{+36}_{-31}$} &  {\small 619.6/618}&   99.74$\%$  \\ 
\noalign {\smallskip}                       
% diskline
\hline
\noalign {\smallskip}                       
{\small PL + {\sc diskline$^{(a)}$}} & {\small 1.58$\pm0.04$} & {\small 6.40$\pm$0.05} &    \multicolumn{1}{c}{ -- }  & {\small $<$22}& {\small 3.9$\pm$1.7} & 107$\pm$46 & {\small 619.3/618} & 99.77$\%$ \\
\noalign {\smallskip}                       
% laor
\hline
\noalign {\smallskip}                       
{\small PL  + {\sc laor}$^{(b)}$} & {\small 1.58$\pm0.04$} & {\small 6.40$^{+0.05}_{-0.10}$} &   \multicolumn{1}{c}{ -- }   & {\small $<$22}& {\small 4.4$\pm$2.1} & 119$^{+59}_{-55}$ & {\small 621.7/618} & 99.30$\%$ \\      
\noalign {\smallskip}                       
\hline
\hline
\end{tabular}
\end{center}
\label{tab:line}
\end{table*}

\subsubsection{The soft spectrum: 0.2--2\,keV}\label{sec:ss}

First we fit the 0.2--2\,keV energy range with a
single absorbed power-law, and found a good fit with
  $N^{\rm int}_{\rm H}$= 2.1$\pm$0.7 $\times$ 10$^{20}$\,cm$^{-2}$ 
 and a photon index of $\Gamma$=1.96$\pm$0.05
 ($\chi^{2}$/d.o.f.=364.9/355). 
 This value is compatible within the error bars with $\Gamma$=2.2$\pm$0.2  
inferred from a {\sl ROSAT PSPC} observation performed in 1992
\citep{deKool94}. 
This value is marginally smaller than the one found in general in
radio-quiet quasars (\citealt{P04a}; $<$$\Gamma_{\rm 0.3-2\,keV}$$>$=2.43 with a
dispersion of 0.35). 
The inferred 0.2--2\,keV unabsorbed flux is 
 2.9$\times$ 10$^{-12}$\,erg\,cm$^{-2}$\,s$^{-1}$ 
(L=1.1 $\times$ 10$^{44}$\,erg\,s$^{-1}$). 
However, adding a narrow ($\sigma$=10\,eV) Gaussian line,
 we found a significant improvement 
to the fit, with $\chi^{2}$/d.o.f.=339.4/353 (as a comparison, Monte
Carlo simulations, as described 
in $\S$\ref{sec:trs}, give $>$99.9$\%$). 
The energy of the line is 556$\pm$15\,eV and its equivalent width ($EW$) 
is 15.9$^{+4.5}_{-5.6}$\,eV. These parameters are consistent 
with the position of the He-like \ion{O}{vii} triplet between 561--574 eV. 
We also fit the data assuming two absorption edges, and we 
 again found a significant improvement to the fit, with 
 $\chi^{2}$/d.o.f.=339.9/351 (F-test $>$99.99$\%$). The energies (and 
optical depths, $\tau$) of the two edges are  
$E$=0.714$^{+0.030}_{-0.032}$\,keV ($\tau$=0.16$\pm$0.06) 
and $E$=0.953$^{+0.055}_{-0.140}$\,keV ($\tau$=0.10$\pm$0.05) respectively.
The energy of the first edge is consistent with the rest-frame energy 
of the \ion{O}{vii} edge (0.740\,keV). The second edge is not compatible
with the rest-frame of the \ion{O}{viii} edge (0.871\,keV) 
or of \ion{Ne}{ix} (1.196\,keV), but could be one of these ions with a 
velocity shift or perhaps a blend of the two edges. 
It also could be due the presence of absorption lines, for example, the 
Fe~L   complex
seen in absorption in several AGN (e.g., \citealt{Sako01}). 
 Unfortunately the present RGS data do not have good enough S/N
 to determine the physical nature of this feature. 
 However, as we will discuss below, the whole average spectrum
could be equally well explained either by  
X-ray ionized reflection on the accretion disk surface,  by a warm
absorber-emitter plus power-law model, or by a smeared
absorption/emission from a high velocity outflow.

%%%% subsection
\subsubsection{The Fe\,K${\alpha}$ complex}

As shown in Fig.~\ref{fig:spectrum}, there is the presence of a
Fe\,K$\alpha$ complex between 5.4 and 6.4\,keV in the quasar frame. 
Therefore, we fit the 2--12\,keV energy range with an absorbed
power-law continuum and three possible profiles for 
the iron line; either (i) a simple Gaussian
({\sc zgauss} in {\sc xspec}), (ii) 
 a line profile corresponding to emission from a relativistic accretion disk 
around a non-rotating black hole ({\sc diskline}; \citealt{Fabian89}), or (iii)
 emission from around a maximally rotating black hole ({\sc laor}; \citealt{Laor91}). 
The spectral fits were improved by the addition of the iron line, 
when modeled either by a 
Gaussian line or by a relativistic accretion disk-line profile, 
with a significance greater than or equal 
to 99.3$\%$ from an F-test (see Table~\ref{tab:line}).
 The line, if modeled by a Gaussian, appears to be more likely narrow 
($\sigma<$0.11\,keV) and produced by neutral to moderately ionized iron, 
with $E$=6.34$\pm$0.05\,keV.
If the line is due to a relativistic accretion disk, the inclination 
of the disk is constrained to be less than about 20$^{\circ}$. 
The line could be associated with X-ray fluorescence from the
molecular torus. 
However, we will see below with time-resolved spectroscopy that
 the situation is more complex, with the possible presence of two
 non-simultaneous transient iron lines at $\sim$5.5\,ks and
 $\sim$6.3--6.4\,keV appearing during the first 5.5\,ks and last 16\,ks 
of the observation, respectively.

%%%% subsection 
\subsubsection{The broad-band spectrum: 0.2--12\,keV}\label{sec:bbs}

The 0.2--12\,keV energy range is well-fit by the combination 
of a blackbody and a power-law component. 
We find $N^{\rm int}_{\rm H}<$ 2.4$^{+0.5}_{-0.4}\times$10$^{20}$\,cm$^{-2}$,
 $kT_{\rm bb}$=149$^{+20}_{-17}$~eV 
and $\Gamma$=1.62$^{+0.28}_{-0.14}$ ($\chi^{2}$/d.o.f.=1008.7/977). 
 A broken power-law model also gave a good fit, with 
$N^{\rm int}_{\rm H}$= 2.9$^{+1.0}_{-1.1}\times$10$^{20}$\,cm$^{-2}$, 
$\Gamma_{\rm soft}$=2.04$^{+0.08}_{-0.09}$, 
$\Gamma_{\rm hard}$=1.59$^{+0.03}_{-0.04}$, 
and $E_{\rm break}$=1.64$^{+0.26}_{-0.13}$~keV  ($\chi^{2}$/d.o.f.=995.8/977). 

In order to obtain a more physical representation of the soft excess,
we have also tested multi-temperature disc models, which may be
expected if the soft X-ray excess originates via thermal emission from
the inner accretion 
disc in PG\,1416$-$129. The {\sc diskbb} (non-relativistic) and {\sc diskpn} 
(relativistic) models were used, together with a power-law to 
model the hard X-ray emission above 2 keV. 
Equally good fits are obtained for both models. 
 For the first model, we found $kT_{\rm diskbb}$=211$^{+33}_{-26}$\,eV 
and $\Gamma$=1.60$\pm$0.03 ($\chi^{2}$/d.o.f.=996.0/977), 
 and similar parameters were found for the second model:
$kT_{\rm diskpn}$= 200$^{+32}_{-24}$\,eV 
and $\Gamma$=1.60$\pm$0.03 ($\chi^{2}$/d.o.f.=996.3/977). 
 The inner disc temperatures obtained through either of these models 
 are much larger than the temperature at 3\,$R_{\rm S}$ ($R_{\rm S}$=2$GM/c^{2}$) 
expected from a standard steady state $\alpha$ thin 
accretion disc, i.e. 8--10\,eV, {\citep{SS73}}, 
assuming a black hole mass of about 3--6$\times$10$^{8}$\,$M_{\sun}$. 
This result is consistent with the temperature found in other
low-redshift AGN, with a 100--200\,eV temperature component being
too hot to be the direct thermal disk emission as
 inferred from the soft X-ray spectra \citep{GD04,P04a,Pi05}. 

 The whole spectrum could be explained with a warm
absorber/emitter combined with a power law continuum. 
Therefore, we used a grid of models generated by the
{\sc xstar} (version 21l) photoionization code \citep{Kallman96} 
to fit the EPIC$-$pn data. A one zone model was adopted 
(covering the range of parameters 1$<$log\,$\xi<$5 and
21$<$log\,$N^{\rm warm}_{\rm H}<$24).  
Solar elemental abundances were assumed.  
A low turbulence velocity of 200\,km\,s$^{-1}$ was used, 
 and the covering factor was fixed to unity. 
The warm absorber column density ($N^{\rm warm}_{\rm H}$) 
was fixed to the lowest value in the {\sc xstar} grid, 10$^{21}$\,cm$^{-2}$, 
while the ionization parameter ($\xi$) was left free. 
A good fit was obtained ($\chi^{2}$/d.o.f.=1077.6/975) 
with $N_{\rm H}^{\rm int}<$1.6$\times$10$^{19}$\,cm$^{-2}$, 
$\Gamma$=1.69$\pm$0.02, 
log\,$\xi$=1.90$^{+0.03}_{-0.10}$ erg\,cm\,s$^{-1}$. 
In this model, the photoionized emission from the warm absorber represents about 7$\%$ of the
total flux in the 0.2--2\,keV energy range. 

\begin{figure}[!Hb]
\resizebox{\hsize}{!}{\rotatebox{-90}{\includegraphics{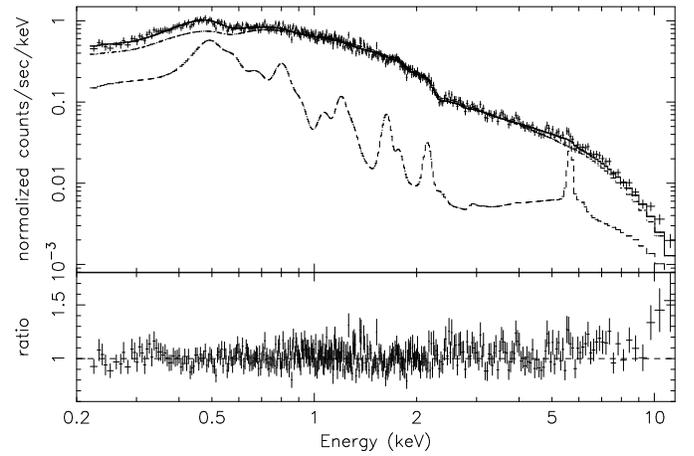}}}
\caption{The time-averaged pn spectrum of PG\,1416$-$129 (in the
  observed frame). 
A relativistically blurred photo-ionized disc model ({\sc reflion}:
\citealt{RF05}) has been fit, and which provides a good representation of the
whole energy band. A primary power-law component is taken into
account in the model. Both component models are displayed without blurring in the
  upper panel. See text for the values of the fit parameters. 
For presentation only, the data have been re-binned into groups of 10 bins,
after group of a minimum of 20 counts per bin is used for the fit.
}
\label{fig:reflion}
\end{figure}

We then tested the possibility that the whole spectrum could be explained
by X-ray ionized reflection on the accretion disk surface 
({\sc reflion} model in {\sc XSPEC}; \citealt{RF05}).
\cite{RF05} have shown that this model can explain  
 the overall spectrum from 0.2--12\,keV in several AGN, without invoking 
ad-hoc components such as a blackbody (see
also \citealt{Crummy05}). 
 One signature of ionized reflection is the soft excess emission which
 occurs in the 0.2--2\,keV band due to lines and bremsstrahlung emission from the
 hot surface layers. It creates a bump in the relativistically blurred spectrum
 which, when convolved with the {\sl XMM-Newton} pn response matrix,
 is well fit by a blackbody of temperature 150\,eV.
The iron abundance (A$_{\rm Fe}$) is from \citet{MMc83} in this
model. The {\sc reflion} model covers a large range of parameters:
30 $\leq$ $\xi$ $\leq$ 10000 erg\,cm\,s$^{-1}$ (ionization parameter), 
0.1 $\leq$ A$_{\rm Fe}$  $\leq$ 10.0 (relative iron abundance), 
and 1.0 $\leq$ $\Gamma$ $\leq$ 3.0. 
We also took into account the relativistic effect due to relativistic
motion in the inner part of the accretion disc, by blurring the 
spectrum with a Laor line profile ({\sc kdblur}; \citealt{RF05,Crummy05}). 
 In the blurred reflection model, the physical lower limit of the
  inner radius ($R_{\rm in}$) was set at 1.235\,$R_{\rm g}$,
i.e. corresponding to the innermost stable orbit around a maximally
rotating black hole.  
We fixed the disc emissivity index to $-$3. 
We obtained a very good fit ($\chi^{2}$/d.o.f.=980.9/974) to the data 
with the following parameters: 
$N^{\rm int}_{\rm H}$=$3.6\pm1.3\times10^{20}$\,cm$^{-2}$, 
$R_{\rm in}$=3.1$^{+1.6}_{-1.0}$\,$R_{\rm g}$ ($R_{\rm g}$=$GM/c^{2}$), 
$\theta$=21.1$^{+10.2}_{-18.8}$$^{\circ}$, 
$A_{\rm Fe}$=0.92$^{+0.45}_{-0.31}$, 
$\Gamma$=1.78$\pm$0.04, and
 $\xi$=38.0$^{+12.8}_{-5.1}$\,erg\,cm$^{-1}$\,s$^{-1}$ 
(see Fig.~\ref{fig:reflion}). 
No additional absorption or emission features in the soft band were
required. As shown in Fig.~\ref{fig:reflion}, the 
  reflection emission is significant in the very soft part of the
  spectrum, especially near 0.4--0.5\,keV (observed frame), where a feature at
  556\,eV has been detected using a simple power-law plus a narrow Gaussian (see $\S$\ref{sec:ss}). 
The fraction of the reflection emission to the total emission is
 only about 21$\%$ in the 0.2--12\,keV 
(similar values are found in the 0.2--2\,keV and 2--10\,keV energy
ranges). 
The inner radius found here is compatible with a rotating black hole. 
The fit parameters found here are compatible to those determined in the large
sample of type I AGN from \cite{Crummy05}. 
The power-law index is in the low range value of the sample 
but not as low than that found for \object{Mrk 359} ($\Gamma$=1.49$\pm$0.06), 
which is an extreme NLS1.

 An alternative possibility is that the soft X-ray excess is an
artifact of soft X-ray absorption and emission produced by a high
velocity outflow, as shown by \cite{GD04} and \cite{Schurch06}, 
who model the X-ray spectra of other 
PG quasars observed by {\sl XMM-Newton}. 
Therefore we fit the data with smeared ionized absorption and
emission, using the same grid of models generated by the
{\sc xstar} photoionization code used for the warm
absorber-emitter model used above. We fixed 
$N^{\rm warm}_{\rm H}$ to 10$^{21}$\,cm$^{-2}$.  
We obtained a very good fit to the data ($\chi^{2}$/d.o.f.=990.8/975), 
with a smearing velocity of 0.18$\pm$0.02\,c, 
$N^{\rm int}_{\rm H}$=9.0$^{+2.9}_{-3.9}\times$10$^{19}$\,cm$^{-2}$, 
 log\,$\xi$=2.31$^{+0.15}_{-0.32}$\,erg\,cm$^{-1}$\,s$^{-1}$, and
 $\Gamma$=1.59$^{+0.02}_{-0.03}$.\\

 As we will see in the next section, an additional physical
  process, such as hot spots due to X-ray flares, is required to
  explain the two possible transient Fe\,K$\alpha$ lines observed
  ($\S$\ref{sec:trs}).\\  

\begin{figure}[!Ht]
\resizebox{\hsize}{!}{\rotatebox{0}{\includegraphics{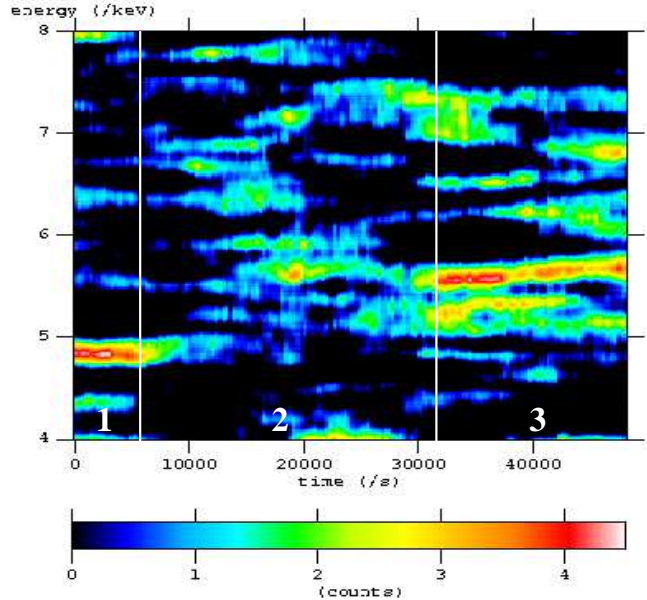}}}
\caption{The signal-to-noise deviations (in $\sigma$) of the
        Fe\,K$\alpha$ line above the
        power-law continuum in the energy-time plane. The energy axis is given in the observed
        frame. The colour-scale represents excess signal/noise
        in the line counts above the fitted continuum. 
        Energy and time are oversampled by a
        factor of 10. Data are top-hat smoothed by 10 ks in time and
        Gaussian smoothed with  $FWHM$=0.14\,keV in energy. The
        regions 1--3 correspond to the time selections used for the
        time-resolved spectroscopic analysis. See text for explanation.}
\label{fig:SN}
\end{figure}

%--------------------- TABLE ---------------------
\begin{table}[!Ht]
\caption{Best-fit spectral parameters for the three pn sub-spectra
 in the 2--10\,keV energy range. Column (1): number of the
 corresponding sub-spectrum (the extraction time range is reported on
 Fig.~\ref{fig:SN}). Column (2): model used. PL:
 power-law; zgauss: Gaussian line ($\sigma$ fixed to 10\,eV); AN: narrow
 diskline annulus
($\Delta R$=1\,$R_{\rm g}$; other diskline parameters are discussed in
the text, see $\S$\ref{sec:trs}).
Column (3): power-law index. Column (4): unabsorbed 2-10\,keV flux
($\times$10$^{-12}$\,erg\,cm$^{-1}$\,s$^{-1}$). Column (5): line
parameters. }
\begin{tabular}{lllllll}
\hline
\hline
\noalign {\smallskip}                       
 n$^{\rm o}$   &{\small Model} &  \multicolumn{1}{c}{\small $\Gamma$} &
 $F^{\rm unabs}_{2-10}$     &  \multicolumn{1}{c}{$E_{\rm line}$ (keV)}  & $\chi^{2}$/d.o.f.  \\
      &    &    &   & \multicolumn{1}{c}{$EW$ (eV)} \\
\noalign {\smallskip}                       
\hline
\noalign {\smallskip}                       
 1 & {\small PL }       &  1.61$\pm$0.11   &  3.2$\pm$0.5  & \multicolumn{1}{c}{ -- }  &  83.8/97  \\
 & \multicolumn{1}{c}{PL+zgauss} & 1.64$\pm$0.11 & \multicolumn{1}{c}{
   -- } & 5.47$\pm$0.04 & 71.0/95   \\
      &    &    &   & 194$\pm$89 \\
\noalign {\smallskip}                       
\hline
\noalign {\smallskip}                       
 2 & {\small PL }   &  1.51$\pm$0.05   &  3.3$\pm$0.2 &
 \multicolumn{1}{c}{ -- }  & 98.8/97  \\
\noalign {\smallskip}                       
\hline
\noalign {\smallskip}                       
 3 & {\small PL }  & 1.61$\pm$0.06   &  3.1$\pm$0.2  & \multicolumn{1}{c}{ -- }  &  117.9/97  \\
 & \multicolumn{1}{l}{PL+zgauss} & 1.62$\pm$0.06 & \multicolumn{1}{c}{
   -- } & 6.35$\pm$0.05  &  102.5/95 \\
      &    &    &   & 128$\pm$53 \\
 & \multicolumn{1}{l}{PL+AN} & 1.64$\pm$0.06 &
 \multicolumn{1}{c}{ -- } & 6.40 (f)  &  95.0/93 \\
      &    &    &   & 287$\pm$100 \\
\noalign {\smallskip}                       
\hline
\hline
\end{tabular}
\label{tab:subspectra}
\end{table}

%%%%% SECTION

\subsection{  {\sl XMM-Newton} time-resolved spectroscopy}\label{sec:trs}

To test possible rapid energy shifts of the iron line, we have created 
X-ray intensity maps in the energy-time plane using the pn data. 
Photons from the source cell were accumulated in pixels in the
energy-time plane. The pixel distribution was smoothed in energy by the
instrumental resolution, using a Gaussian of 140\,eV
(appropriate for the single plus double events with the latest
calibration), and smoothed in time using a top-hat function of width 10\,ks. Each
time-slice was background-corrected by subtracting a time-dependent
background spectrum measured in an off-source region on the same
detector chip as the source. The source continuum was modeled as an
absorbed power-law, with variable amplitude and slope but time-invariant
absorption column density. This continuum was subtracted, leaving positive and
 negative residuals that comprise noise plus any emission or
 absorption components on top of the continuum. 
More information about this method can be found in \cite{Tu06}.
The ``signal-to-noise'' (S/N) map presented in Fig.\ref{fig:SN}
 is the ratio of the fluctuation amplitude to the calculated noise. 

\begin{figure}[!Ht]
%\begin{tabular}{cc}
\resizebox{\hsize}{!}{\rotatebox{-90}{\includegraphics{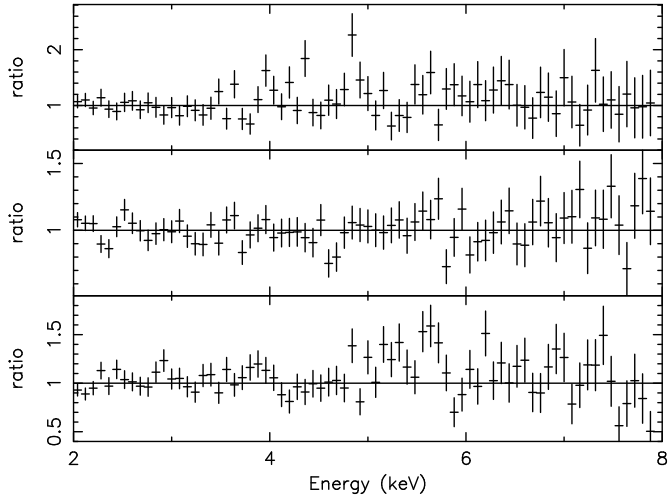}}}
%\end{tabular}
\caption{The three pn time-resolved spectroscopy spectra  (in the
  observed frame) obtained by
  splitting the average spectra according to the time regions shown
in Fig.\,3. The three sub-spectra have been fit with a
power-law continuum model. Here, 5.67\,keV in the observed frame corresponds 
to 6.4\,keV in the rest frame. Notice the residuals present at 4.8\,keV and 
5.6\,keV in segments 1 (upper panel) and 3 (lower panel), respectively.}
\label{fig:subspectra}
\end{figure}
The highest S/N observed is about 4.5. We observe the presence of an excess 
 with S/N $\geq$3.5 at $\sim$5.5\,keV in the quasar frame at the
 beginning of this observation. Then this feature disappears. At the
 end of this observation, there is the presence of an excess near
 6.3\,keV (in the quasar frame) and then at lower significance
 (S/N $\leq$3.5) 
a hint of a shift to 6.4\,keV of the centroid of the line. 
In order to characterize these features, which are most probably
associated with Fe\,K$\alpha$ line emission, we have split the observation
into three sub-spectra (the corresponding time ranges are illustrated on 
Fig.~\ref{fig:SN}): \\
- 1. This corresponds to the presence of the 5.5\,keV feature with
$S/N\geq$ 3.5 (from t=0\,ks to t=5.5\,ks);\\
- 2. This corresponds to the absence of any significant long-lived
excess feature (from t=5.5\,ks to 32\,ks);  \\
- 3. This corresponds to the appearance and presence of the feature
near 6.3--6.4\,keV.

\begin{figure}[!Ht]
\resizebox{\hsize}{!}{\rotatebox{-90}{\includegraphics{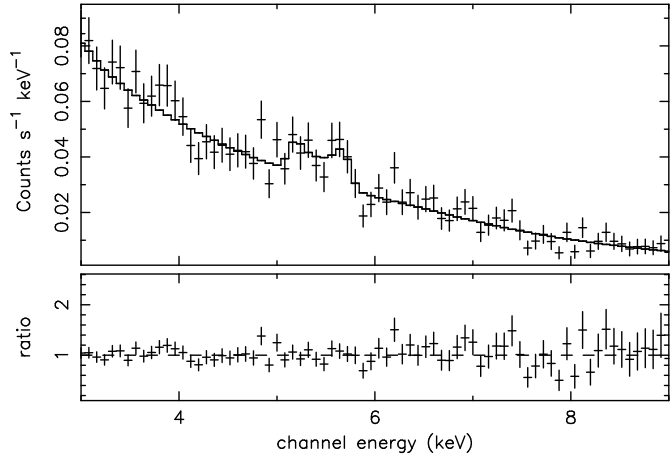}}}
\caption{The pn sub-spectrum 3 of PG\,1416$-$129 (in the observed frame). 
A power law continuum plus a {\sc diskline} model have been fit. A
narrow annulus is assumed with $\Delta$R=1\,$R_{\rm g}$. See text for
the best fit parameter values. }
\label{fig:annuli}
\end{figure}

First, we rebinned the sub-spectra, grouping every 80\,eV (16 channels).
This was chosen to sample the resolution of the pn detector (i.e. about 
2 spectral bins per $FWHM$ resolution element). We then fit them with a
power-law continuum over the 2--10\,keV energy range.
The parameter fits are reported in Table~\ref{tab:subspectra} and 
the spectral fits are displayed in Figure~\ref{fig:subspectra}. 
For sub-spectra 1 and 3, an emission feature is observed.
 For sub-spectrum 1, the addition of a narrow 
Gaussian line (width fixed to 10\,eV) with a centroid energy of
$\sim$5.5\,keV improved the fit. Allowing the line width to
vary, we found an upper limit of 94\,eV. 
For sub-spectrum 3, the addition of a narrow Gaussian (width fixed to 10\,eV) 
with a centroid energy of about 6.35\,keV also improved the fit. 
  If we allow the line width to be a free parameter, the upper limit is
  rather large ($\sigma\leq$0.4\,keV) due to the presence of the red
  wing-like feature seen below 6.4\,keV (see Fig.~\ref{fig:subspectra}).    
A narrow disc annulus at 6.4\,keV ($\Delta
R$=1\,$R_{\rm g}$, emissivity index fixed to $-$2) provides a superior fit.  
The best fit parameters are $\Gamma$=1.64$\pm$0.06, 
 $R_{\rm in}$=34$^{+5}_{-8}\,R_{\rm g}$, $\theta$=17$\pm$3$^{\circ}$ 
and $EW$=287$\pm$100\,eV (Fig.~\ref{fig:annuli}). Letting the value of
the accretion disc outer radius free to vary, we find an upper limit
of 84\,$R_{\rm g}$, corresponding to a $\Delta\,R$ of the annulus 
value between 1 and 58\,$R_{\rm g}$ (taking into account the lower
limit on $R_{\rm in}$). 

As discussed by \cite{Protassov02} (see also \citealt{P04b}), the F-test
can overestimate the true detection significance of a line
if there was no previous expectation of the line.
One must account for the number of resolution elements
$N$ over the bandpass of interest. The probability
of detecting a feature at any energy in this range 
is obtained using $(P_1)^N$, where $P_1$ is the probability of finding 
a feature at an expected line energy (equal to one minus the
null hypothesis probability).
Assuming 21 resolution elements between 4--7 keV,
the significance probabilities for the 5.5 and 6.3\,keV lines
become 98.1$\%$ and 99.2$\%$, respectively.
We also carried out a more rigorous test of the significance 
of the lines using Monte Carlo simulations to assess the 
likelihood that the lines could be an artefact of photon noise 
(e.g., \citealt{P04b}). For our null hypothesis, we assumed 
that the spectrum is simply an absorbed power-law continuum, 
with the same parameters as the absorbed power-law model fit 
to the real data. We used the \textsc{xspec fakeit none} command to create 
a fake spectrum, and then, to account for the uncertainty in the null 
hypothesis model itself, we refit the null hypothesis model and 
ran \textsc{fakeit none} a second time on the ``re-fit" null hypothesis
model. Both \textsc{fakeit none} commands were run with the photon 
statistics appropriate for a 5.5 ksec exposure (segment 1) or 
16 ksec exposure (segment 3).  We grouped each faked 
spectrum, rebinning every 80 eV. We then added a narrow 
($\sigma$ = 10 eV) Gaussian to the fit for the 5.5 keV line, 
or a diskline annulus, with the values of the inner and outer radii, 
emissivity index, and inclination angle fixed to that of the 
observed spectrum, for the 6.3\,keV line.
We searched over 4.0 to 7.0 keV (observed 
frame) in increments of 0.1 keV, fitting separately each time to 
ensure that the lowest value of $\chi^2$ was found. We
compared the minimum value of $\chi^2$ obtained 
with the corresponding $\chi^2$ of the null hypothesis fit.
This process was repeated 1000 times, yielding
a cumulative frequency distribution of the simulated $\Delta\chi^2$
expected for a blind line search. The inferred 
probabilities to reject the null hypothesis are 98.8$\%$ and 99.7$\%$
for the 5.5\,keV and 6.3\,keV lines, respectively. 
 However, since we have divided the whole observation into three
  sub-spectra, the null hypothesis probabilities have to be multiplied
  by a factor of three, leading to 96.4$\%$ and 99.1$\%$ for the
  detection confidence level of these transient lines. 
While the detection of the 5.4\,keV feature is statistically marginal, 
it should be noted that the 6.3\,keV feature is observed at an energy where a 
iron K$\alpha$ diskline would be expected a priori.\\
The three sub-spectra display no significant flux variations:
$F^{\rm unabs}_{2-10}$ is consistent with being constant.
As far as the continuum photon index is concerned, sub-spectra 1 and 3
have consistent values, but sub-spectrum 2 is slightly
flatter; the photon index when the diskline annulus model is applied
to sub-spectrum 3 is not consistent at 90$\%$ confidence
with that obtained in sub-spectrum 2, where no emission line was found.
However, we can use Monte Carlo simulations to
quantify the expected spread in the measured photon index, given
an assumed intrinsic constant photon index.
Specifically, the null hypothesis model in the
Monte Carlo procedure above assumed a single value
for the intrinsic photon index; using those same simulated
spectra, we find a standard deviation in the
measured photon index of 0.165 (due to the moderate signal/noise).
This is higher than the maximum $\Delta$$\Gamma$ of 0.13 between the
best-fit photon indices for sub-spectra 3 and 2.
Therefore, there is no evidence
for the appearance of the line  (or lack thereof)
to be correlated with values of the photon index.

\section{Summary and discussion}

In this section, we summarize the main results of our analysis of 
the July 2004 {\sl XMM-Newton}/EPIC observation of PG\,1416$-$129. \\

\noindent $\bullet$ A X-ray flat RQQ object.\\
With the present 2004 {\sl XMM-Newton} observation, we found 
that its 2--5\,keV power-law index
($\Gamma\sim$1.6) is similar to values found for radio-loud objects 
 ($\Gamma$=1.74 with a standard deviation of 0.03; \citealt{P04a}),
but by far much harder than the value commonly seen in radio-quiet quasars:
 $\Gamma$=1.90 (with a standard deviation of 0.27) and 
 $\Gamma$=2.37 (with a standard deviation of 0.11) for 
Broad Line quasars and Narrow Line quasars, respectively (\citealt{P04a}). 
 This result confirms the behavior seen in previous observations, 
with the exception of the 1994 {\sl ASCA}
observation, as
reported in Table~\ref{tab:flux}.\\

 \noindent $\bullet$ Is PG1416$-$129 a NLS1 ? \\
PG\,1416$-$129 could be associated with the NLS1 class of objects, given the
value of $FWHM$(H$\beta$) observed in 2000 by
\cite{Sulentic00}.  \\
NLS1s are defined according to the following characteristics: \\
$-$ $FWHM$(H$\beta$) less than 2000\,km\,s$^{-1}$; strong optical
 \ion{Fe}{ii} multiplets; 
 $[\ion{O}{iii}]\lambda$5007\AA/H$\beta_{\rm n}\leq$ 3
 \citep{OP85,Goodrich89}. The 2000 optical spectrum of PG1416$-$129, however,
showed only a weak \ion{Fe}{ii} multiplet, with
$[\ion{O}{iii}]\lambda$5007\AA/H$\beta\sim$13.9
\citep{Sulentic00}. \\
$-$ steep, soft X-ray excess \citep{Pu92,Bo96};
 rapid soft/hard X-ray variability \citep{Leighly99}. 
During the 2004 {\sl XMM-Newton}
observation, however, neither significant X-ray variability nor a steep soft
excess were observed.  \\
$-$ less massive black holes with higher
Eddington ratios, suggesting that they might be in the early stage of
AGN evolution \citep{Grupe96,Mathur00,BZ03}. As reported in the
introduction, however, this object has a black hole mass of about 
3$-$6$\times$10$^{8}$\,$M_{\odot}$, which is higher than the mass found in
typical NLS1s. Furthermore, its accretion rate with respect to
Eddington is only 0.12. \\
Therefore it appears that PG\,1416$-$129 is not a 
 genuine NLS1 and instead the narrow H$\beta$ width is caused by the 
variations in the broad component of this line. \\

\noindent $\bullet$ The soft X-ray spectrum. \\
In addition to its flat 2-10\,keV photon index, this object shows
a much weaker soft excess than generally observed for Broad-Line
quasars and Narrow-Line quasars. Some additional features seem to be
present in the 0.2--2\,keV energy range. For example, the addition of
a Gaussian emission line at 556$\pm$15\,eV improved the fit very
significantly, with an F-test significance of $>$99.99$\%$. However, we have
obtained a fit with the same statistical quality by adding two absorption
edges. Unfortunately the present RGS data do not have  sufficient
signal to noise ratio in order to determine the physical nature of
this feature, if real.  
The whole average spectrum is very well fit by a model with 
X-ray ionized reflection on the accretion disk surface, 
 with no additional soft X-ray absorption and/or emission features
  required. 
However other models, such as absorption 
and/or re-emission from a high velocity 
outflow (or simple photoionized emission from distant matter)
could also explain the weak soft excess 
below 1\,keV, as discussed in detail for other PG quasars by \citet{GD04}. \\

\noindent $\bullet$ Discovery of two possible transient
  Fe\,K$\alpha$ lines\\  
In the time-averaged spectrum, a Fe\,K${\alpha}$ complex is seen
between 5.5 and 6.4\,keV, with significant residuals at about 6.4\,keV.
If the line is due to a relativistic accretion disk, the inclination 
of the disk is constrained to be less than about 20$^{\circ}$. 
If the line is variable, this would rule out X-ray fluorescence from distant 
matter such as the molecular torus. 
Indeed, time-resolved spectroscopy has revealed that the emission complex 
is more likely due to the presence of two transient
iron lines.
The first iron line has a redshifted energy centered at
$\sim$5.5\,keV and was present during the first $\sim$5.5\,ks
($EW\sim$167\,eV) of this observation with a 96.4$\%$
confidence level according to Monte-Carlo simulations (see
$\S$\ref{sec:trs}); after 5.5 ksec, 
it disappears sharply. The second, more significant 
transient iron line appears about 32\,ks after the beginning of this
observation with a centroid energy at $\sim$6.3\,keV ($EW\sim$250\,eV) 
and a duration of at least $\sim$16\,ks with a 99.1$\%$ confidence
level according to Monte-Carlo simulations. There is 
 a hint of a slight shift in its centroid energy 
to $\sim$6.4\,keV during the last 10\,ks of the observation. 
 Other narrow, highly-redshifted iron lines
($\sim$4.5--6\,keV) in AGN have already been reported: e.g., 
\object{NGC\,3516} (Turner et al. 2002), 
\object{ESO\,113-G010} \citep{P04b},
\object{Mrk\,766} \citep{JTurner04}, 
%\object{IC 4329A} \citep{McKernan04},  
\object{AX J0447-0627} \citep{DellaCeca05}, 
and \object{PG\,1425+267} \citep{MF06}.
Localized spots or narrow annuli which occur on the surface of an 
accretion disk following its illumination by flares have been proposed 
to explain these features (e.g., \citealp{Nayakshin00,Tu02,Dovciak04,Pe05}).
Such lines may arise in the accretion disk, 
sporadically illuminated by ``hotspots,''
as regions of magnetic reconnection illuminate very
small areas of the accretion flow. 
Presumably the hotspot co-rotates with the disk.
 While the small line width implies that the emitting region must
  be small and 
detected only during a fraction of the whole orbit,  
large spots, meanwhile, would produce broader features and they would
be prone to 
rapid destruction. Here the features appear to be narrow, however
their widths are not strongly constrained. 
 Indeed, an annulus of up to 58\,$R_{\rm g}$ in size is allowed in
  segment 3 of the observation. 
 If the transient lines are due to hotspots 
  co-rotating with the disk, a lower limit of 16\,ks on the lifetime
  of the transient 6.3\,keV iron line feature would 
correspond to at least 80$\%$ and 2$\%$ of the whole orbit at
 $R$=1.235\,$R_{\rm g}$ and $R$=20\,$R_{\rm g}$, respectively (assuming a
 black hole mass of 3 $\times$10$^{8}$\,$M_{\sun}$).
 The large $EW$ of both transient features are consistent with 
 the X-ray emission being located in a relatively compact region, e.g., 
from X-ray flares. 
 More complex geometries of the X-ray emitting region can be
  considered, e.g., spiral waves propagating across the accretion
  disc (e.g., \citealt{Karas01,Fukumura04}), 
and density inhomogeneities, 
which can indicate turbulence driven by magnetorotational instability
\citep{Balbus91}, or photon bubble instabilities \citep{Arons92}. 
As demonstrated by \cite{Ballantyne04,Ballantyne05}. Photon bubble
instabilities (which result on time-varying density homogeneities on
scales smaller than the disk thickness) can explain rapid variability
of emission lines, with $EW$ variations as large as $\sim$100\,eV that 
are independent of the illuminating continuum, as observed
 during the present observation. 
In addition,  scenarios involving non-flat accretion disks, such as
warped/corrugated accretion disks (e.g., \citealt{Fabian02a, Miller05}), 
can be invoked to explain such features. However, as pointed out by
\cite{Fabian02a}, even a small corrugation of the accretion disc  
 may result in $\Gamma>$2 and a strong reflection component in the
observed spectrum, as observed in several NLS1 objects (e.g,
\object{1H 0707$-$495}). In the case of the present observation of
PG\,1416$-$129  such a  scenario appears unlikely since neither a steep
spectrum ($\Gamma\sim$1.6) nor a strong reflection component 
(the fraction of the 0.2--12\,keV reflection emission is only about
21$\%$ of the total emission, see $\S$\ref{sec:bbs}) are observed.  

\begin{acknowledgements}
The {\sl XMM-Newton} project is an ESA Science Mission with instruments
and contributions directly funded by ESA Member States and the
USA (NASA). The {\sl XMM-Newton} project is supported by the
Bundesministerium f\"ur Wirtschaft und Technologie/Deutsches Zentrum
f\"ur Luft- und Raumfahrt (BMWI/DLR, FKZ 50 OX 0001), the Max-Planck
Society and the Heidenhain-Stiftung.
We thank the anonymous referee for fruitful comments and suggestions.
D.P. acknowledges grant support from an MPE fellowship.
\end{acknowledgements}

\appendix
\section{RXTE data}\label{sec:rxte}
PG~1416$-$129 was observed by {\it RXTE} on 1998 Aug 21 
from 06:25--10:42 UT, 1998 Aug 22 from 06:54--08:06 UT, 
and 1998 Aug 23 from 01:37--07:32 UT, for a total of 41.1 
ksec. Data were taken using {\it RXTE}'s proportional
counter array (PCA; \citealp{Swank98}), which consists of five 
identical collimated proportional counter units (PCUs). 
Data were collected from PCUs 0, 1, and 2, and reduced 
using standard extraction methods and {\sc FTOOLS~v5.3.1} 
software. Data were rejected if they were gathered less 
than 10$^{\circ}$ from the Earth's limb, if they were obtained 
within 30~min after the satellite's passage through
the South Atlantic Anomaly, if {\sc ELECTRON0}~$>$~0.1, 
or if the satellite's pointing offset was greater
than 0$\fdg$02.\\
As the PCA has no simultaneous background monitoring capability, 
background data were estimated by using {\sc pcabackest~v3.0} 
to generate model files based on the particle-induced background, 
SAA activity, and the diffuse X-ray background. The 'L7-240' 
background models appropriate for faint sources were used. 
This background subtraction is the dominant source of systematic
error in {\sl RXTE} AGN monitoring data (e.g., \citealt{EN99}). 
Counts were extracted only from the topmost PCU layer to maximize 
the signal-to-noise ratio. This reduction yielded 
a total of 22.0 ksec of good exposure time.\\
Response matrices were generated using {\sc pcarsp v.8.0}.
For the purposes of spectral analysis, data below 3.0 keV were discarded
in order to disregard PCA calibration uncertainties
below this energy. The source was not well detected by the PCA
above $\sim$14 keV; photons above this energy were discarded. 

Fitting these RXTE data with an absorbed ($N_{\rm H}^{\rm Gal}$) 
power-law continuum, we found a good fit with
$\Gamma$=1.17$\pm$0.15 ($\chi^{2}$/d.o.f.=7.6/25). 
The inferred 2--10\,keV luminosity is 1.0 $\times$ 10$^{44}$\,erg\,s$^{-1}$. 

\bibliographystyle{aa}
\bibliography{biblio}

\end{document}